\documentclass[aps,twocolumn,prd]{revtex4-1}
\usepackage{graphicx}
\usepackage{amssymb}
\usepackage{slashed}
\newcommand{\be}{\begin{eqnarray}}
\newcommand{\ee}{\end{eqnarray}}
\usepackage[colorlinks]{hyperref}
\usepackage{cancel}

\usepackage{amsmath,amssymb}
\usepackage{mathtools}
\usepackage{float}
\usepackage{color}
\usepackage{leftidx}
\usepackage{comment}
\usepackage{booktabs}
\hypersetup{
    colorlinks=black,
    citecolor=blue,
    filecolor=magenta,      
    urlcolor=red,
}
\begin{document}

\title{Accessing unpolarized and linearly polarized gluon TMDs through quarkonium production }

\author{Asmita Mukherjee and Sangem Rajesh}

\altaffiliation{This proceeding is based on a talk delivered at 22$^\mathrm{nd}$ International Spin 
Symposium, 2016, UIUC.}
\email{rajeshphy@phy.iitb.ac.in}

\affiliation{ Department of Physics,
Indian Institute of Technology Bombay, Mumbai-400076,
India.}
\date{\today}

\begin{abstract}
We present the study of accessing unpolarized and linearly polarized gluon TMDs in $J/\psi$ and $\Upsilon(1\mathrm{S})$ 
production in unpolarized proton-proton collision at LHC, RHIC and AFTER energies. Non-relativistic QCD based color 
octet model (COM) is used for estimating quarkonium production rates within transverse momentum dependent factorization formalism.
A comparison is drawn between the experimental data and the transverse momentum distribution of quarkonium obtained in COM and color 
evaporation model.
\end{abstract}

\maketitle

\section{Introduction}\label{sec1}
Amongst the eight leading twist-2 transverse momentum dependent parton distribution functions (TMDs) 
\cite{Mulders:2000sh,Angeles-Martinez:2015sea,Meissner:2007rx}, $f_1^{g}(x,{\bf 
k}_\perp)$ and $h_1^{\perp g}(x,{\bf k}_\perp)$ are the only two TMDs which describe the dynamics of 
gluons inside an unpolarized hadron while $f_1^{g}$ and $h_1^{\perp g}$ represent the density of unpolarized and 
linearly polarized gluons inside an unpolarized  hadron respectively. TMDs have been receiving paramount interest 
in both theoretically  and experimentally as they provide the  3-dimensional structure and spin information of the 
nucleon. TMDs depend on both longitudinal momentum fraction ($x$) and intrinsic transverse momentum ($k_\perp$) of the 
parton whereas usual collinear parton distribution functions (PDFs) depend only on $x$.
Gauge links are required to define the gauge invariant operator definition of TMDs and are process dependent.
\par
In general, linearly polarized gluons can be present even at tree level inside an unpolarized hadron 
\cite{Mulders:2000sh} provided that the 
gluons carry transverse momentum w.r.t parent hadron. The associated density function of linearly polarized gluons, 
$h_1^{\perp g}$ is a T-even (time-reversal even) distribution and is also even in the transverse momentum. TMDs are 
nonperturbative objects and have to be extracted from experiments.
Drell-Yan (DY) and semi-inclusive deep inelastic scattering (SIDIS) are the two processes which provide the 
experimental data related to the TMDs \cite{Melis:2014pna}. In these processes, the intrinsic transverse momentum 
($k_\perp$) has an 
imprint on the experimentally measurable quantities, for instance, azimuthal asymmetries and transverse momentum 
($p_T$) distribution of the final hadron. Hence, these quantities are very sensitive to the TMDs.
However, $h_1^{\perp g}$ and  even $f_1^{ g}$ 
have not been extracted yet.
Gluon Sivers function ($f_1^{\perp g}$) \cite{Sivers:1989cc} generates single spin asymmetry in scattering processes 
like $ep^\uparrow$ and 
$pp^\uparrow$. In order to understand asymmetries fully, one should have complete knowledge about unpolarized TMDs 
since $f_1^{g}$ sits in the denominator of the asymmetry expression \cite{Melis:2014pna}.
Therefore, the extraction of $f_1^{ g}$ and $h_1^{\perp g}$  functions are of prime importance. In order to 
probe $h_1^{\perp g}$, several processes have been proposed theoretically. Linear gluon polarization can be determined 
by measuring azimuthal asymmetry in heavy quark pair and dijet production in SIDIS \cite{Pisano:2013cya}, 
$\Upsilon+jet$ \cite{Dunnen:2014eta} and $\gamma\gamma$ \cite{Qiu:2011ai}
in $pp$ collision at LHC.  $h_1^{\perp g}$ can also be accessed through the  cross section of 
Higgs-boson 
\cite{Boer:2011kf,Boer:2013fca,Echevarria:2015uaa,Boer:2014tka}, Higgs+jet  \cite{Boer:2014lka} and C-even (charge 
conjugation even) quarkonium production \cite{Boer:2012bt}.\par
In this proceeding contribution, we discuss the $J/\psi$ and $\Upsilon(1\mathrm{S})$ production in  unpolarized 
proton-proton 
collision  to show that the quarkonium production is also a promising channel to 
extract both  $f_1^{ g}$ and 
$h_1^{\perp g}$.
Details of our work can be found in \cite{Mukherjee:2016cjw,Mukherjee:2015smo}.
We estimate the quarkonium production rates using color octet model (COM) \cite{Mukherjee:2016cjw} 
within transverse momentum 
dependent (TMD) \cite{jcollins} framework and draw a comparison between the results with color evaporation model (CEM) 
\cite{Mukherjee:2015smo} and experimental data. COM, color singlet model (CSM) and CEM are 
the three important models for quarkonium production, which are successful at different energies. Generally, two scales 
are involved in quarkonium production \cite{Amundson:1995em,Amundson:1996qr,Bodwin:1994jh}. The first one is related to 
the production of heavy quark pair with momentum of 
order $M$ (heavy quark mass) which is called short distance factor. This short distance factor can be calculated in 
order $\alpha_s(M)$ using perturbation theory. The second one is the binding of quarkonium bound state which 
is taking place at scale of order $\Lambda_{\mathrm{QCD}}$. This is a nonperturbative process and is denoted with long 
distance matrix elements  (LDME) in factorization expression. The hadronization information is encoded in the LDME 
which are usually extracted by fitting data. The non-relativistic Quantum chromodynamics (NRQCD) effective field theory 
\cite{Bodwin:1994jh} separates the short 
distance and long 
distance factors systematically. In COM \cite{Bodwin:1992ye}, the initially produced heavy quark pair can be either in 
color singlet or 
octet state.  
\section{$J/\psi$ and $\Upsilon(1\mathrm{S})$ production in COM}
We consider unpolarized proton-proton collision process for quarkonium production
$i.e.,$ $p+p\rightarrow J/\psi~\mathrm{or}~\Upsilon(1\mathrm{S})+~X$. Proton is rich of gluons at high 
energy, hence we consider the leading order (LO) gluon-gluon fusion channel for quarkonium production. Assuming that 
the TMD factorization holds good, the differential cross section is given by \cite{Mukherjee:2016cjw}
\begin{equation}\label{cross1}
 \begin{aligned} 
 {d\sigma}={}&\int dx_{a} dx_{b}  d^{2}{\bf k}_{\perp a} d^2{\bf k}_{\perp b}
 \Phi^{\mu\nu}_g(x_{a},{\bf k}_{\perp a})\\
& \times\Phi_{g\mu\nu}(x_{b},{\bf k}_{\perp b})
 {d\sigma^{J/\psi(\Upsilon)}},
\end{aligned}
\end{equation}
where, $\Phi^{\mu\nu}_g$ is the gluon-gluon correlator of unpolarized spin-$\frac12$ hadron, which can be 
further parametrized in terms of leading twist-2 TMDs as the following \cite{Mulders:2000sh} 
\be \label{corre}
\begin{aligned}
\Phi^{\mu\nu}_g(x,{\bf k}_{\perp})={}&-\frac{1}{2x}\Big\{g^{\mu\nu}_Tf^g_1(x,{\bf 
k}_{\perp}^2)-\Big(\frac{k^{\mu}_{\perp}k^{\nu}_{\perp}}
{M^2_h}\\
& +g^{\mu\nu}_T\frac{{\bf k}^2_{\perp}}{2M_h^2}\Big)
 h^{\perp g}_1(x,{\bf k}_{\perp}^2)\Big\}.
\end{aligned}
\ee
Here $f_1^{ g}$ and $h_1^{\perp g}$ are the  unpolarized and linearly polarized gluon TMDs respectively. $M_h$ is the 
proton mass.
The ${d\sigma^{J/\psi(\Upsilon)}}$ in Eq.\eqref{cross1} is the  partonic differential cross section of  $gg\rightarrow 
Q\bar{Q}[\leftidx{^{2S+1}}{L}{^{(a)}_J}]$ channel. Using NRQCD, the partonic differential cross section can 
be 
factorized as follows \cite{Fleming:1995id,Bodwin:1994jh}
\be\label{e1}
d\sigma^{J/\psi(\Upsilon)}=\sum_n d{\hat{\sigma}}[gg\rightarrow Q\bar{Q}(n)]
\langle 0\mid \mathcal{O}^{J/\psi(\Upsilon)}_n\mid 0\rangle
\ee
The first term in the right hand side of Eq.\eqref{e1} was given in \cite{Mukherjee:2016cjw} that describes 
the production of heavy quark and anti-quark pair in 
a definite quantum state and it can be calculated in order $\alpha_s$. Spin, orbital angular momentum and color quantum 
numbers are denoted with $n$. After forming the heavy quark pair, its  quantum 
numbers will be  readjusted to form a color 
singlet quarkonium state by emitting or absorbing soft gluons. This process is absorbed in  $\langle 0\mid 
\mathcal{O}^{J/\psi(\Upsilon)}_n\mid 0\rangle$ (LDME) which is nonperturbative. All possible 
configurations of heavy quark pair in different quantum states are taken into account for quarkonium 
production which is represented with summation over $n$ in Eq.\eqref{e1}. 
In line with  Ref. \cite{Fleming:1995id,Cooper:2004qe}, we consider only the color octet states 
$\leftidx{^{1}}{S}{_0}$, $\leftidx{^{3}}{P}{_0}$ and $\leftidx{^{3}}{P}{_2}$ which have
 dominant  contribution in charmonium and bottomonium production. The LDME numerical values of these color 
octet states are extracted in Ref.\cite{Ma:2014mri,Chao:2012iv,Sharma:2012dy}, which are tabulated in 
\cite{Mukherjee:2016cjw}. After integrating w.r.t $x_a$, $x_b$ and ${\bf k}_{\perp b}$ in Eq.(\ref{cross1}) 
and  following the steps in Ref. \cite{Mukherjee:2016cjw}, one can obtain the differential cross section as 
\be\label{d1}
\frac{d\sigma^{ff+hh}}{dyd^2{\bf p}_T}=\frac{d\sigma^{ff}}{dyd^2{\bf p}_T}+\frac{d\sigma^{hh}}{dyd^2{\bf 
p}_T},
\ee
where
\begin{equation}
\begin{aligned}
\frac{d\sigma^{ff}}{dyd^2{\bf p}_T}={} &\frac{C_n}{s} \int d^2{\bf k}_{\perp a}
f_1^g(x_{a},{\bf k}_{\perp a}^2)
 f_1^g(x_{b},{\bf k}_{\perp b}^2),
\end{aligned}
\end{equation}

\begin{equation}
\begin{aligned}
\frac{d\sigma^{hh}}{dyd^2{\bf p}_T}={} &\frac{C_n}{s} \int 
d^2{\bf k}_{\perp a} w
 h_1^{\perp g}(x_{a},{\bf k}_{\perp a}^2)h_1^{\perp g}(x_{b},{\bf k}_{\perp b}^2),
\end{aligned}
\end{equation}

$w=\frac{1}{2M_h^4}\left[\left({\bf k}_{\perp a}.{\bf k}_{\perp b}\right)^2-\frac{1}{2}
{\bf k}_{\perp a}^2{\bf k}_{\perp b}^2 \right]$ and ${\bf k}_{\perp b}={\bf p}_{T}-{\bf 
k}_{\perp a}$. The definition of $C_n$ is given in Eq.(6) of Ref. \cite{Mukherjee:2016cjw}.    
Here $p_T$ and $y$ are the transverse momentum and rapidity of the quarkonium.

\section{Evolution of TMDs}  

As per Ref. \cite{Boer:2012bt}, we assume that the unpolarized and linearly polarized gluon TMDs follow 
the Gaussian form. In Gaussian parametrization, TMDs are factorized into product of collinear PDFs  times 
exponential factor which is a function of only $k_\perp$ and Gaussian width.
\be \label{unp}
 f_1^{g}(x,{\bf k}^2_{\perp })=f_1^{g}(x,Q^2)\frac{1}{\pi \langle k^2_{\perp }\rangle}
 e^{-{\bf k}^2_{\perp }/\langle k^2_{\perp }\rangle},
\ee
\be\label{hg}
h_1^{\perp g}(x,{\bf k}^2_{\perp })=\frac{M^2_hf_1^g(x,Q^2)}{\pi\langle k^2_{\perp 
}\rangle^2}\frac{2(1-r)}{r}e^{1-
 {\bf k}^2_{\perp }\frac{1}{r\langle k^2_{\perp }\rangle}},
\ee
where, $f_1^{g}(x,Q^2)$  is the collinear PDF which follows the DGLAP evolution equation and 
$r=2/3$ and 1/3 
\cite{Boer:2012bt} values are taken for numerical estimation. The Gaussian widths are $\langle 
k^2_{\perp 
}\rangle=0.25$ GeV$^2$ and 1 GeV$^2$ \cite{Boer:2012bt}.
In model-I, we  do not take any upper limit for $k_{\perp a }$ integration. An upper limit 
$k_{\mathrm{max}}=\sqrt{\langle {k}^2_{\perp }\rangle}$ \cite{Anselmino:2008sga} is considered for $k_{\perp 
a}$ integration in model-II. The analytical expressions of differential cross sections for model-I and 
model-II are given in Sec-(III) \cite{Mukherjee:2016cjw}.
As pointed out in Ref.\cite{Melis:2014pna}, in order to explain high $p_T$ spectrum one has to consider the full TMD 
evolution approach which was derived in impact parameter space ($b_\perp$). The Fourier transformations of 
gluon-gluon correlator in $b_\perp$ and $k_\perp$ space are 
\be\label{et1}
\Phi(x,{\bf b}_\perp)=\int d^2{\bf k}_\perp e^{-i{\bf k}_{\perp}.{\bf b}_{\perp}}\Phi(x,{\bf k}_\perp),
\ee
\be\label{et2}
\Phi(x,{\bf k}_\perp)=\frac{1}{(2\pi)^2}\int d^2{\bf b}_\perp 
e^{i{\bf k}_{\perp}.{\bf b}_{\perp}}\Phi(x,{\bf b}_\perp).
\ee

The gluon correlator in $b_\perp$ space is given by \cite{Boer:2014tka}
\be\label{et3}
\begin{aligned}
\Phi^g(x,{\bf b}_\perp)={}&\frac{1}{2x}\Big\{g^{\mu\nu}_Tf^g_1(x,{\bf b}_{\perp}^2)-
\Big(\frac{2b^{\mu}_{\perp}b^{\nu}_{\perp}}{b^2_{\perp}}\\
&-g^{\mu\nu}_T\Big)h^{\perp g}_1(x,{\bf b}_{\perp}^2)\Big\}.
\end{aligned}
\ee
In TMD evolution approach, TMDs depend on both renormalization scale $\mu$ and auxiliary scale $\zeta$ which 
was introduced to regularize the rapidity divergences. Renormalization group (RG) and Collins-Soper (CS) 
equations are obtained  by taking scale evolution w.r.t the scales $\mu$ and $\zeta$. After solving these 
equations one obtains the TMD evolution expressions of TMDs in $b_\perp$ space 
\cite{Aybat:2011zv,Aybat:2011ta,jcollins}. The  differential cross section expressions of Eq.\eqref{d1} in 
TMD evolution  approach are given by \cite{Mukherjee:2016cjw}
 \begin{equation}\label{evoleq1}
\begin{aligned}
 \frac{d^2\sigma^{ff}}{dydp^2_T}={} &\frac{C_n}{2s}
 \int_0^{\infty}b_{\perp} db_{\perp}J_0(p_Tb_{\perp})
 f^g_1(x_a,c/b_{\ast})\\
 &\times f_1^g(x_b,c/b_{\ast}) R_{\mathrm{pert}}R_{\mathrm{NP}},
  \end{aligned}
\end{equation}

and
\begin{equation}\label{evoleq2}
\begin{aligned}
 \frac{d^2\sigma^{hh}}{dydp^2_T}={}&\frac{C_n C_A^2}{2s\pi^2}
 \int_0^{\infty}b_{\perp} db_{\perp}J_0(p_Tb_{\perp})\alpha_s^2(c/b_{\ast})\\
 &\times\int_{x_a}^1\frac{dx_1}{x_1}\left(\frac{x_1}{x_a}-1\right)f^g_1(x_1,c/b_{\ast})\\
&\times\int_{x_b}^1\frac{dx_2}{x_2}\left(\frac{x_2}{x_b}-1\right)
f_1^g(x_2,c/b_{\ast}) R_{\mathrm{pert}}R_{\mathrm{NP}}
 \end{aligned}
\end{equation}
where $R_{\mathrm{pert}}$ and $R_{\mathrm{NP}}$ are the perturbative and nonperturbative parts of 
the 
evolution kernel.

 \be
 R_{\mathrm{pert}}=
\mathrm{exp}\Bigg\{{-2\int_{c/b_{\ast}}^{Q}\frac{d\mu}{\mu}\left(A\log\left(\frac{Q^2}
 {\mu^2}\right)+B\right)}\Bigg\}\nonumber
 \ee
 \be
 R_{\mathrm{NP}}=
 \mathrm{exp}\Bigg\{-\Big[0.184\log\frac{Q}{2Q_0}
 +0.332\Big]b_{\perp}^2\Bigg\} \nonumber
 \ee
Here $A$ and $B$ are the anomalous dimensions of the evolution kernel and TMDs respectively and these have 
perturbative expansion \cite{Mukherjee:2016cjw}. We used the $b_\ast$ prescription to avoid 
the Landau poles by freezing the scale as 
$b_{\ast}(b_{\perp})=\frac{b_{\perp}}{\sqrt{1+\left(\frac{b_{\perp}}{b_{\mathrm{max}}}\right)^2}}$. In the 
nonperturbative regime where $b_\perp$ is very large, the evolution kernel cannot be 
calculated using perturbation theory. Hence, the evolution kernel in this regime is modeled as 
$R_{\mathrm{NP}}$ \cite{Aybat:2011zv}. We have considered the same nonperturbative factor $R_{\mathrm{NP}}$  for 
both unpolarized and linearly polarized gluon TMDs.
\begin{figure}[h]

\normalfont{(a)}\includegraphics[width=\linewidth,height=0.8\linewidth]{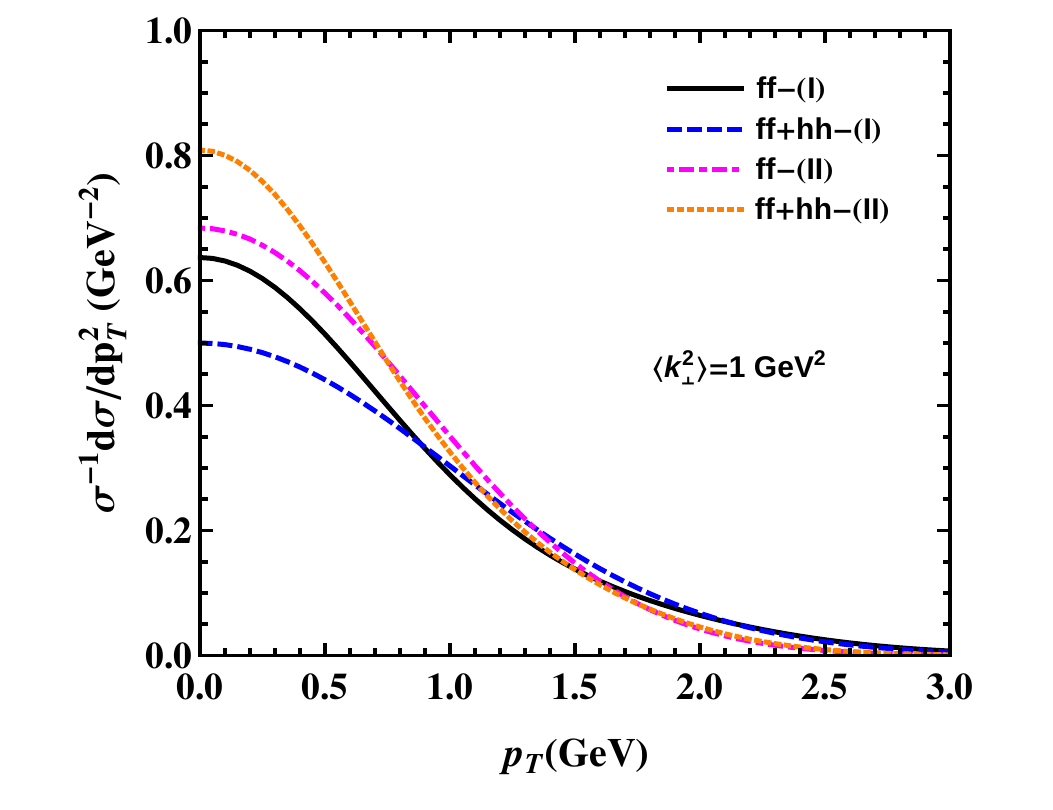}

\normalfont{(b)}\includegraphics[width=\linewidth,height=0.8\linewidth]{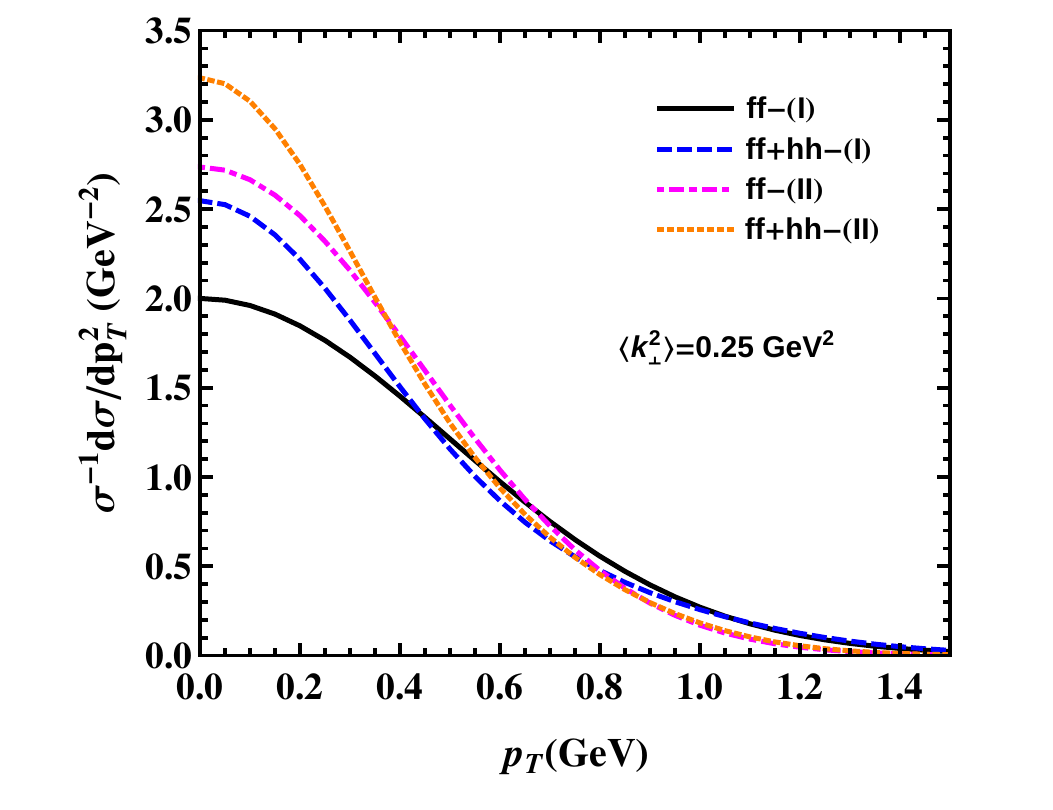}

\caption{\label{fig1}(color online) Differential cross section (normalized) of 
 $J/\psi$  and $\Upsilon(1\text{S})$ production in $pp\rightarrow J/\psi(\Upsilon(1\text{S}))+X$  
at LHCb ($\sqrt{s}=7$ TeV),  RHIC ($\sqrt{s}=500$ GeV) and AFTER ($\sqrt{s}=115$ GeV) energies
using DGLAP evolution approach  for (a) ${\langle {k}^2_{\perp }\rangle}=1$ GeV$^2$   and (b) 
 ${\langle {k}^2_{\perp }\rangle}=0.25$ GeV$^2$  at  $r=\frac{2}{3}$  .
The solid (ff-(I)) and dot dashed (ff-(II)) lines are obtained by considering
unpolarized gluons  in Model-I and Model-II respectively.
The dashed (ff+hh-(I)) and tiny dashed (ff+hh-(II)) lines are obtained by  taking 
into account unpolarized gluons plus linearly polarized gluons in Model-I and Model-II
respectively. See the text for ranges of rapidity integration \cite{Mukherjee:2016cjw}.}
\end{figure}	
\begin{figure}[h]
\normalfont{(a)}\includegraphics[width=0.85\linewidth,height=0.8\linewidth]{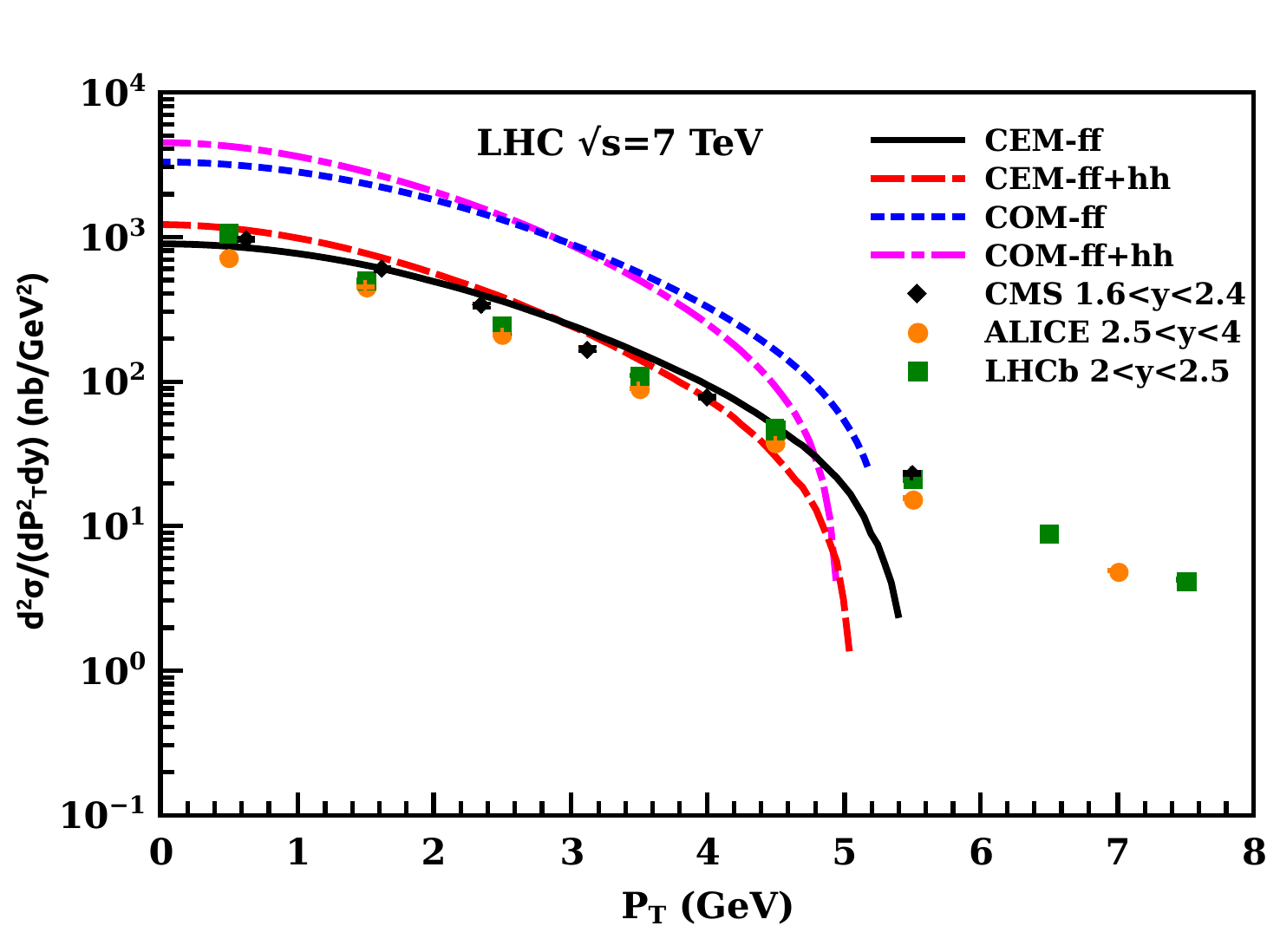}
\normalfont{(b)}\includegraphics[width=0.89\linewidth,height=0.8\linewidth]{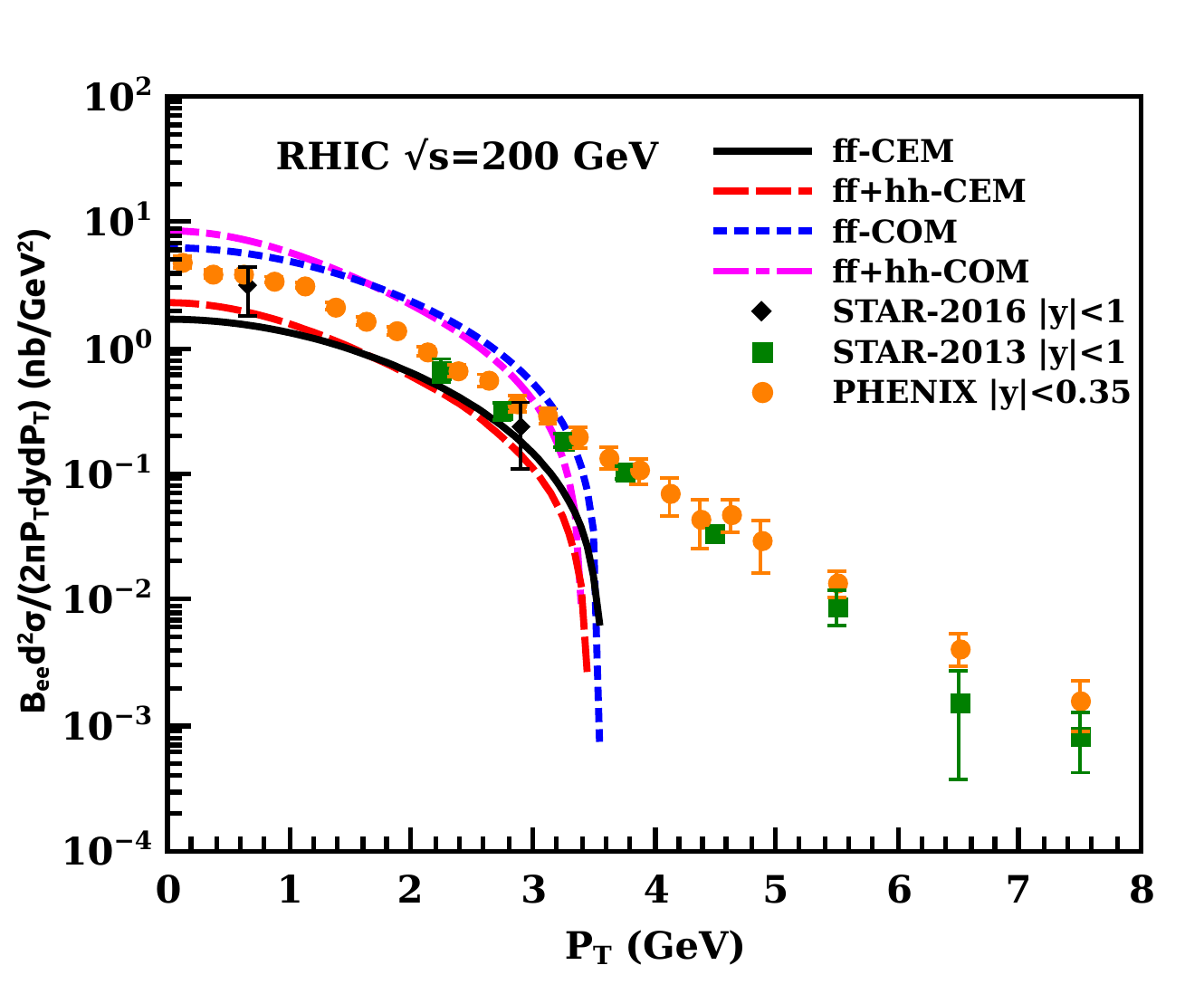}
\caption{\label{fig2}(color online). Differential cross section of $J/\psi$  at (a) LHCb 
($\sqrt{s}=7$ TeV) 
 and  (b) RHIC ($\sqrt{s}=200$ GeV)   as function of $p_T$
in $pp\rightarrow J/\psi+X$ using TMD evolution approach.
Data are taken from \cite{Aaij:2011jh,Khachatryan:2010yr,Abelev:2014qha} and 
\cite{Adamczyk:2012ey,Adamczyk:2016dhc,Adare:2009js}
for LHC and RHIC respectively. The 
rapidity in the range $2.0<y<2.5$  and  $-0.35<y<0.35$ is chosen for LHCb and RHIC energies 
respectively \cite{Mukherjee:2016cjw}.}
\end{figure}
\begin{figure}[h]
\includegraphics[width=0.87\linewidth,height=0.8\linewidth]{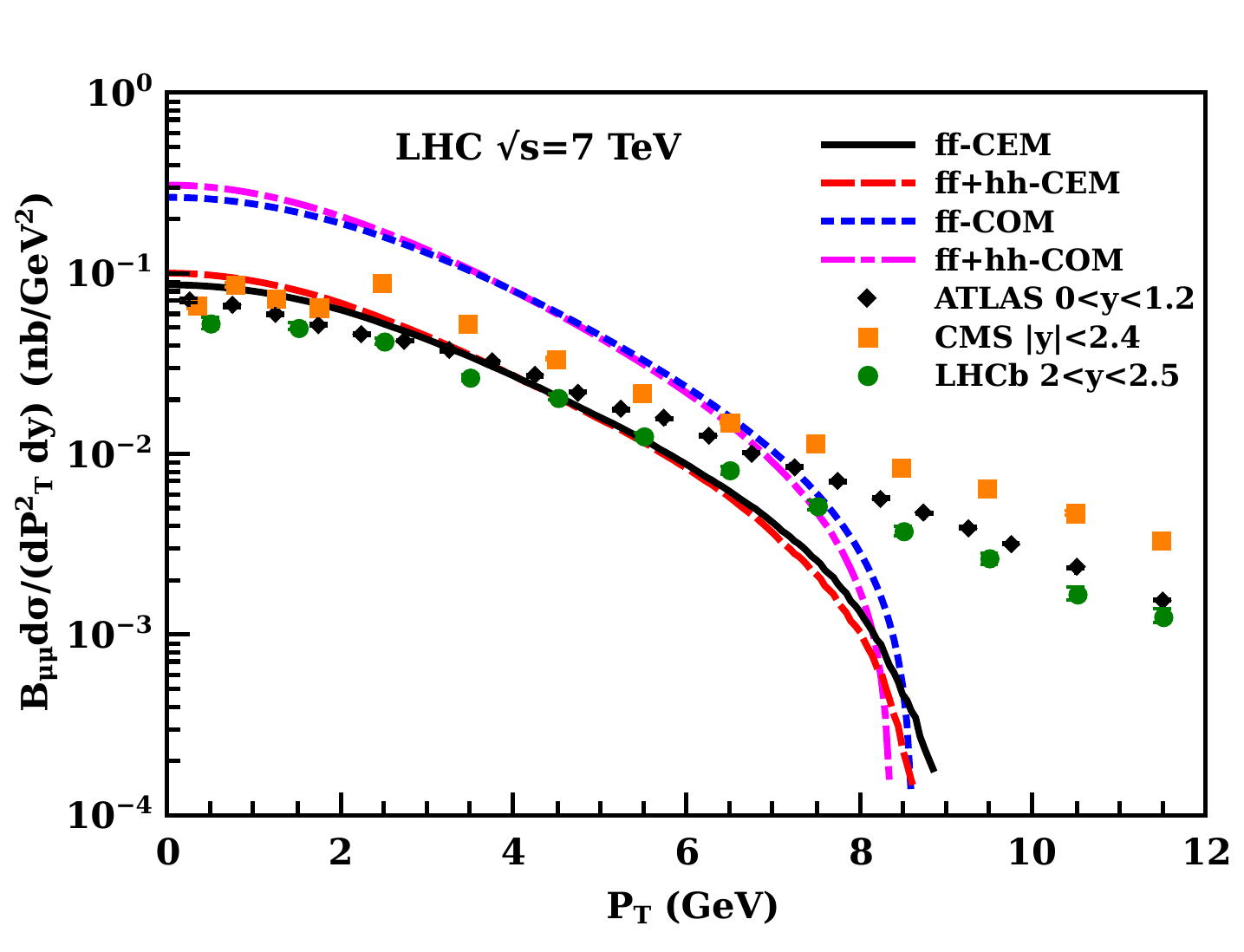}
\caption{\label{fig3}(color online). Differential cross section of $\Upsilon(1\text{S})$ at LHCb 
($\sqrt{s}=7$ TeV) as function of $p_T$
in $pp\rightarrow \Upsilon(1\text{S})+X$ using TMD evolution approach.
Data are taken from \cite{LHCb:2012aa,Chatrchyan:2013yna,Aad:2012dlq}. The rapidity in the range 
$2.0<y<2.5$ 
is chosen \cite{Mukherjee:2016cjw}.}
\end{figure}
\section{Numerical Results}

We calculated the transverse momentum ($p_T$) distribution of  $J/\psi$ and 
$\Upsilon(\mathrm{1S})$  in unpolarized proton-proton collision at LHC ($\sqrt{s}=7$ TeV ), RHIC 
($\sqrt{s}=500$ GeV ) and AFTER ($\sqrt{s}=115$ GeV ) energies. Quarkonium production rates are estimated 
using NRQCD version COM within TMD factorization framework. Color octet states such as 
$\leftidx{^{1}}{S}{_0}$, 
$\leftidx{^{3}}{P}{_0}$ and $\leftidx{^{3}}{P}{_2}$ of initially produced heavy quark pair  are taken into 
account for quarkonium production. The masses of $J/\psi$ and   $\Upsilon(\mathrm{1S})$ are considered 3.096 
and 9.398 GeV respectively. $m_c=1.5$ GeV and 
$m_b=4.8$ GeV are taken for charm and bottom quark masses respectively. MSTW2008 \cite{Martin:2009iq} is used for gluon 
PDFs. $Q=M$ (quarkonium mass) is considered for scale of the gluon PDFs in DGLAP evolution. Quarkonium 
$p_T$ distribution is obtained by integrating rapidity in the range of
$y\in[2.0,4.5]$, $y\in[-3.0,3.0]$ and  $y\in[-0.5,0.5]$ for LHCb, RHIC and AFTER respectively. The convention 
in the figures as follows. \textquotedblleft ff\textquotedblright and \textquotedblleft ff+hh\textquotedblright 
represent the quarkonium distribution obtained by taking into 
account only unpolarized gluons and linearly polarized plus unpolarized gluons respectively. \par
\figurename{\ref{fig1}} represents the $p_T$ spectrum of $J/\psi$ and $\Upsilon(\mathrm{1S})$ 
which is estimated in COM.
In \figurename{\ref{fig1}}, the  cross section differential in $p_T$ is normalized with total cross section 
as a result we obtain the $p_T$ spectrum which is independent of center of mass energy and quarkonium mass.
The obtained $p_T$ spectrum in DGLAP evolution approach in model-I and model-II are compared in 
\figurename{\ref{fig1}} at $r=2/3$.
The quarkonium $p_T$ spectrum has been modulated  significantly  by taking into consideration  of
linearly polarized gluons along with the unpolarized gluons in the scattering process.  The effect of 
linearly  polarized gluons is more in model-II compared to model-I. In \figurename{\ref{fig2}},
the estimated $p_T$ spectrum of $J/\psi$ in TMD evolution approach at LHCb and RHIC energies in COM 
and CEM are compared with 
data. Experimental data is taken from Ref. \cite{Aaij:2011jh,Khachatryan:2010yr,Abelev:2014qha} and Ref. 
\cite{Adamczyk:2012ey,Adamczyk:2016dhc,Adare:2009js} for LHCb and RHIC experiments respectively.
In \figurename{\ref{fig3}},  $p_T$ spectrum of $\Upsilon(\mathrm{1S})$ using TMD evolution approach 
in COM  and CEM is compared with
data \cite{LHCb:2012aa,Chatrchyan:2013yna,Aad:2012dlq}. The production rates are in good accuracy with
data up to low $p_T$ for both  $J/\psi$ and $\Upsilon(\mathrm{1S})$, however, COM is slightly over estimated.
In \figurename{\ref{fig2}} and \figurename{\ref{fig3}}, $\mathrm{B}_{ee}$ (0.0594) and $\mathrm{B}_{\mu\mu}$ 
(0.0248) are the branching ratios of $J/\psi\to e^+e^-$  and $\Upsilon(1\text{S})\to\mu^+\mu^-$  
channels respectively.  $J/\psi$ and $\Upsilon(\mathrm{1S})$ states can be produced from higher 
mass excited states. However, we have considered only the direct production of quarkonium in this 
article. In general, LO calculation is insufficient to explain full $p_T$ spectrum. It may be 
possible to explain  high $p_T$ spectrum by adding NLO calculation with LO.

\section{Conclusion}
We studied the transverse momentum ($p_T$)  distribution of $J/\psi$ and 
$\Upsilon(\mathrm{1S})$  in 
unpolarized proton-proton collision within TMD factorization formalism. NRQCD based color octet 
model is employed to estimate the quarkonium production rates. The quarkonium $p_T$ spectrum has 
been modulated by the presence of linearly polarized gluons inside unpolarized proton and is 
in good agreement with LHCb and RHIC data. Hence, quarkonium 
production offers a good possibility to probe both unpolarized and linearly polarized gluon TMDs.
\section*{Acknowledgement}
SR acknowledges IIT Bombay and spin symposium organizers for financial support to attend the 
22$^\mathrm{nd}$ International Spin Symposium, 2016, UIUC. 

\bibliographystyle{apsrev} 
\bibliography{reference}

\end{document}